\documentclass[epsfig,twocolumn]{revtex4}
\usepackage{graphicx}
\begin{document}

\title{Hierarchical quantum information splitting}
\author{Xin-Wen Wang$^{1,2,}$\footnote{E-mail Address: xwwang@mail.bnu.edu.cn}, Li-Xin Xia$^{3}$, Zhi-Yong Wang$^{4}$}
\address{$^{1}$Department of Physics,
Hunan University of Science and Engineering, Yongzhou 425100,
China\\
$^{2}$Department of Physics and Applied Optics Beijing Area Major
Laboratory, Beijing Normal University, Beijing 100875, China\\
$^{3}$Department of Physics, Henan University of Science and
Technology, Luoyang, 471003, China\\
$^{4}$ National Laboratory of
Solid State Microstructures, Nanjing University, Nanjing 210093,
China}

\begin{abstract}

\noindent\textbf{Abstract} \\
We present a scheme for asymmetric quantum information splitting,
where a sender distributes asymmetrically a qubit to distant agents
in a network. The asymmetric distribution leads to that the agents
have different powers to reconstruct the sender's qubit. In other
words, the authorities of the agents for getting the quantum secret
are hierarchized. The scheme does not need the agents to get
together and make nonlocal operations. Our scheme can also be
modified to implement controlled teleportation against
uncooperation of part of supervisors.  
\end{abstract}

\maketitle

The combination of information theory and quantum mechanics leads to
the advent of quantum information science \cite{Nielsen}.
Entanglement, one of the most striking features of quantum
mechanics, is the center resource for quantum information
processing. The extensive applications of quantum entanglement
should owe to its nonlocal correlations. One well-known example is
quantum teleportation \cite{70PRL1895,58PRA4394}, which utilizes the
nonlocality of the quantum channel, i.e., bipartite or multipartite
entangled states, to transport an unknown quantum state between two
spatially separated quantum systems. In the original teleportation
protocol of Bennett \emph{et al.} \cite{70PRL1895} the sender
(Alice) and the receiver (Bob) initially share a maximally entangled
state of two particles. Alice then performs a joint measurement on
her particle of the entangled pair and the particle whose state is
to be teleported. With the outcome transmitted to Bob via a
classical channel, he can recover the teleported state by
appropriate local transformations.

Generally, the more particles that can be entangled, the more
clearly nonlocal effects are exhibited \cite{403N515}, and the more
useful the states are for quantum information processing
\cite{86PRL5188}. In addition, the usefulness of entangled states is
usually related to their entanglement properties
\cite{96PRL060502,364PLA7}. Thus exploring and exploiting
multipartite entangled states are very important tasks for the
workers who study quantum information science. It has been
attracting much interest that what classes of multipartite entangled
states are competent for achieving a defined quantum information
processing task and what they can do. Greenberger-Horne-Zeilinger
(GHZ) states \cite{GHZ} is a typical multipartite entangled states.
With the GHZ states Hillery \emph{et al.} \cite{59PRA1829} firstly
introduced the concept of quantum information splitting (QIS), where
a qubit is distributed to two or more distant agents and anyone of
them can reconstruct the original qubit (quantum secret) if and only
if they cooperate. QIS can be considered as a generalization of
teleportation, and was also called open-destination teleportation or
quantum-state sharing in literature \cite{430N54,71PRA033814}.

QIS has extensive applications in the quantum world, such as it
could help us create joint checking accounts containing quantum
money \cite{quantum money}, perform secure distributed quantum
computation \cite{distributed computation}, and so on. Since the
outstanding work of Hillery \emph{et al.}, QIS has been attracting
much attention
\cite{59PRA162,83PRL648,64PRA042311,74PRA054303,78PRA042309}, and a
scheme has already been experimentally realized \cite{430N54}.
However, all of these schemes are focused on the symmetric case
where every participant has the same status, i.e., the same
authority to get the secret. In Ref.~\cite{61PRA042311}, Gottesman
pointed out that a more general QIS scheme should involve the
asymmetry between the powers of the different participants, and
showed that it is possible to construct an access structure that
some subsets of the shares can be combined to reconstruct the secret
quantum state. This case was further studied later
\cite{71PRA012328,72PRA032318}. Their idea is based on theory of
quantum error-correcting codes, and thus the nonlocal operations are
required.

 In this paper, we present a scheme for distributing a qubit to
three distant agents asymmetrically. The asymmetric distribution
leads to that the agents have different powers to reconstruct the
sender's qubit. In other words, the authorities of the agents for
getting the quantum secret are hierarchized. The scheme does not
need the agents to come together and make nonlocal operations.

The quantum channel of our scheme is the four-qubit entangled state,
recently proposed by Yeo and Chua \cite{96PRL060502},
\begin{equation}
\label{X}
 |\chi_{ABCD}\rangle=\frac{1}{\sqrt{2}}\left(|0_A\rangle|\varphi^0_{BCD}\rangle+|1_A\rangle|\varphi^1_{BCD}\rangle\right),
\end{equation}
where
\begin{eqnarray}
  |\varphi^0_{BCD}\rangle&=&\frac{1}{2}(|0_B0_C0_D\rangle-|0_B1_C1_D\rangle\nonumber\\
   && -|1_B0_C1_D\rangle+|1_B1_C0_D\rangle),\nonumber\\
  |\varphi^1_{BCD}\rangle&=&\frac{1}{2}(|0_B0_C1_D\rangle+|0_B1_C0_D\rangle\nonumber\\
  &&+|1_B0_C0_D\rangle+|1_B1_C1_D\rangle).
\end{eqnarray}
The state $|\chi_{ABCD}\rangle$ has many interesting properties and
exhibits more nonlocality than the counterparts of the well-known
GHZ states and $W$ states \cite{96PRL060502,75PRA032332}. In
addition, it can be easily verified that at least two single-qubit
measurements are required in order to completely disentangle
$|\chi_{ABCD}\rangle$. Thus such a state has higher persistency of
entanglement than the GHZ states which can be completely
disentangled by only one local measurement. This may lead to that
our scheme is more robust against decoherence than the scheme of
Ref.~\cite{59PRA1829}.

We consider that Alice, Bob, Charlie, and Diana possess particles
$A$, $B$, $C$, and $D$, respectively. These particles are in the
entangled state $|\chi_{ABCD}\rangle$. Alice has another particle
$S$ which is in the state
\begin{equation}
\label{S}
  |\xi_S\rangle=\frac{1}{\sqrt{1+|\lambda|^2}}(|0_S\rangle+\lambda|1_S\rangle).
\end{equation}
The state of the whole system is
\begin{eqnarray}
\label{whole}
  |\xi_S\rangle|\chi_{ABCD}\rangle &=&
  \frac{1}{\sqrt{2(1+|\lambda|^2)}}(|0_S0_A\rangle|\varphi^0_{BCD}\rangle\nonumber\\
   &&+|0_S1_A\rangle|\varphi^1_{BCD}\rangle)\nonumber\\
   &&+\frac{\lambda}{\sqrt{2(1+|\lambda|^2)}}(|1_S0_A\rangle|\varphi^0_{BCD}\rangle\nonumber\\
   &&+|1_S1_A\rangle|\varphi^1_{BCD}\rangle).
\end{eqnarray}

Alice performs a joint measurement on her two particles $S$ and $A$
with the Bell basis
\begin{eqnarray}
  |\Psi^{\pm}_{SA}\rangle=\frac{1}{\sqrt{2}}(|0_S0_A\rangle \pm |1_S1_A\rangle),\nonumber\\
  |\Phi^{\pm}_{SA}\rangle=\frac{1}{\sqrt{2}}(|0_S1_A\rangle \pm
  |1_S0_A\rangle).
\end{eqnarray}
Then the particles held by Bob, Charlie, and Diana collapse into one
of the following entangled states:
\begin{eqnarray}
  |\psi^{\pm}_{BCD}\rangle=\frac{1}{\sqrt{1+|\lambda|^2}}(|\varphi^0_{BCD}\rangle \pm \lambda|\varphi^1_{BCD}\rangle) , \nonumber\\
  |\phi^{\pm}_{BCD}\rangle=\frac{1}{\sqrt{1+|\lambda|^2}}(|\varphi^1_{BCD}\rangle \pm \lambda|\varphi^0_{BCD}\rangle) .
\end{eqnarray}
The non-cloning theorem \cite{239N802} allows only one particle to
be in the original state of particle $S$, so that anyone of Bob,
Charlie, and Diana, but not all, will recover the original state.

In order to reconstruct Alice's qubit, Bob, Charlie, and Diana need
cooperating. Before they come to an agreement, their single-particle
state-density matrices are
\begin{eqnarray}
  \rho_{B(C)}&=& \frac{1}{2}\left(|0_{B(C)}\rangle\langle 0_{B(C)}|+|1_{B(C)}\rangle\langle
  1_{B(C)}|\right), \nonumber\\
  \rho^{\pm}_{D} &=& \frac{1}{2}(|0_{D}\rangle\langle 0_{D}|+|1_{D}\rangle\langle 1_{D}|) \nonumber\\
  && \pm i\frac{ \mathrm{Im}(\lambda)}{(1+|\lambda|^2)}(|1_{D}\rangle\langle 0_{D}|-|0_{D}\rangle\langle
  1_{D}|),
\end{eqnarray}
where $\rho^+_{D}$ corresponds to Alice's measurement outcomes
$|\Psi^+_{SA}\rangle$ and $|\Phi^-_{SA}\rangle$, and $\rho^-_{D}$
corresponds to $|\Psi^-_{SA}\rangle$ and $|\Phi^+_{SA}\rangle$. It
can be seen that Bob or Charlie knows nothing about the amplitude
and phase of Alice's qubit $S$ without the collaboration of the
other two agents; Diana, however, has partial information about both
the amplitude and phase of qubit $S$ as long as he receives Alice's
Bell-state measurement outcome. This case implies that Alice's qubit
is distributed to Bob, Charlie, and Diana asymmetrically. We shall
show that the asymmetric distribution leads to an interesting
phenomenon: Bob or Charlie can reconstruct Alice's qubit conditioned
on that both of the other two agents cooperate, while Diana has
access to recover the qubit if anyone of the other agents
cooperates.

First, we assume that the three agents agree to let Bob possess the
final qubit. We rewrite $|\psi^{\pm}_{BCD}\rangle$ and
$|\phi^{\pm}_{BCD}\rangle$ as
\begin{eqnarray}
 |\psi^{\pm}_{BCD}\rangle &=& \frac{1}{2\sqrt{1+|\lambda|^2}}[(|0_B\rangle \pm \lambda|1_B\rangle)|0_C0_D\rangle\nonumber\\
     && -(|1_B\rangle \mp \lambda|0_B\rangle)|0_C1_D\rangle +(|1_B\rangle \pm \lambda|0_B\rangle)\nonumber\\
     &&\times|1_C0_D\rangle -(|0_B\rangle \mp \lambda|1_B\rangle)|1_C1_D\rangle],\nonumber\\
 |\phi^{\pm}_{BCD}\rangle &=& \frac{1}{2\sqrt{1+|\lambda|^2}}[(|1_B\rangle \pm \lambda|0_B\rangle)|0_C0_D\rangle\nonumber\\
     && +(|0_B\rangle \mp \lambda|1_B\rangle)|0_C1_D\rangle +(|0_B\rangle \pm \lambda|1_B\rangle)\nonumber\\
     &&\times|1_C0_D\rangle+(|1_B\rangle \mp \lambda|0_B\rangle)|1_C1_D\rangle].
\end{eqnarray}
It can be seen that if Charlie and Diana, respectively, perform a
measurement on their particles with the basis
$\{|0\rangle,|1\rangle\}$ (i.e., along the $z$ direction) and inform
Bob their outcomes, Bob can recover the original state $|\xi\rangle$
on his particle $B$ by appropriate local unitary transformations. In
other words, Bob can reconstruct Alice's qubit if and only if both
Charlie and Diana collaborate with him. In particular, the
transformations that Bob should perform on particle $B$ in order to
recover the state $|\xi\rangle$, up to an overall sign, are
\begin{eqnarray}
 |\Psi^{+}_{SA}\rangle|0_C0_D\rangle\rightarrow I, & & |\Phi^{+}_{SA}\rangle|0_C0_D\rangle\rightarrow \sigma_x, \nonumber\\
 |\Psi^{+}_{SA}\rangle|0_C1_D\rangle\rightarrow \sigma_x\sigma_z, && |\Phi^{+}_{SA}\rangle|0_C1_D\rangle\rightarrow \sigma_z, \nonumber\\
 |\Psi^{+}_{SA}\rangle|1_C0_D\rangle\rightarrow \sigma_x, & &|\Phi^{+}_{SA}\rangle|1_C0_D\rangle\rightarrow I, \nonumber\\
 |\Psi^{+}_{SA}\rangle|1_C1_D\rangle\rightarrow \sigma_z, & &|\Phi^{+}_{SA}\rangle|1_C1_D\rangle\rightarrow \sigma_x\sigma_z, \nonumber\\
 |\Psi^{-}_{SA}\rangle|0_C0_D\rangle\rightarrow \sigma_z, & &|\Phi^{-}_{SA}\rangle|0_C0_D\rangle\rightarrow \sigma_x\sigma_z, \nonumber\\
 |\Psi^{-}_{SA}\rangle|0_C1_D\rangle\rightarrow \sigma_x, && |\Phi^{-}_{SA}\rangle|0_C1_D\rangle\rightarrow I, \nonumber\\
 |\Psi^{-}_{SA}\rangle|1_C0_D\rangle\rightarrow \sigma_x\sigma_z, & &|\Phi^{-}_{SA}\rangle|1_C0_D\rangle\rightarrow \sigma_z, \nonumber\\
 |\Psi^{-}_{SA}\rangle|1_C1_D\rangle\rightarrow I, && |\Phi^{-}_{SA}\rangle|1_C1_D\rangle\rightarrow
 \sigma_x,
\end{eqnarray}
where $I$ is $2\times 2$ identity matrix, $\sigma_x$ and $\sigma_z$
are the usual Pauli matrices. These results are also applicable to
the case where Charlie is deputed to reconstruct Alice's qubit,
because particles $B$ and $C$ are fully symmetrical in the state
$|\chi_{ABCD}\rangle$.

Now, we assume that they agree to let Diana regenerate the state
$|\xi\rangle$. We rewrite $|\psi^{\pm}_{BCD}\rangle$ and
$|\phi^{\pm}_{BCD}\rangle$ as
\begin{eqnarray}
 |\psi^{\pm}_{BCD}\rangle &=& \frac{1}{2\sqrt{1+|\lambda|^2}}[(|0_B0_C\rangle+|1_B1_C\rangle)\nonumber\\
      &&\times(|0_D\rangle \pm
   \lambda|1_D\rangle)-(|0_B1_C\rangle+|1_B0_C\rangle) \nonumber\\
   &&\times(|1_D\rangle \mp \lambda|0_D\rangle)]\nonumber\\
    &=& \frac{1}{\sqrt{2(1+|\lambda|^2)}}[|+_B+_C\rangle(|-_D\rangle \pm \lambda|+_D\rangle)\nonumber\\
     && +|-_B-_C\rangle(|+_D\rangle \mp \lambda|-_D\rangle)],\nonumber\\
 |\phi^{\pm}_{BCD}\rangle &=& \frac{1}{2\sqrt{1+|\lambda|^2}}[(|0_B0_C\rangle+|1_B1_C\rangle)\nonumber\\
    && \times(|1_D\rangle \pm
    \lambda|0_D\rangle)+(|0_B1_C\rangle+|1_B0_C\rangle)\nonumber\\
    &&\times(|0_D\rangle \mp \lambda|1_D\rangle)] \nonumber\\
   &=& \frac{1}{\sqrt{2(1+|\lambda|^2)}}[|+_B+_C\rangle(|+_D\rangle \pm \lambda|-_D\rangle)\nonumber\\
     && -|-_B-_C\rangle(|-_D\rangle \mp \lambda|+_D\rangle)],
\end{eqnarray}
where $|\pm_j\rangle=(|0_j\rangle\pm |1_j\rangle)/\sqrt{2}$
($j=B,C,D$). Then interesting phenomena appear. (1) The
single-particle measurement bases that Bob and Charlie can adopt are
optional, $\{|0\rangle,|1\rangle\}$ or $\{|+\rangle,|-\rangle\}$. In
other words, they can choose anyone of the two sets of bases to
perform projective measurements on their particles in order to
assist Diana to reconstruct Alice's qubit. In the protocol of
Ref.~\cite{59PRA1829}, however,  the case that anyone of the
collaborators adopts the measurement basis $\{|0\rangle,|1\rangle\}$
will result in the failure of recovering the original state of the
sender's particle. (2) If Bob and Charlie choose the measurement
basis $\{|+\rangle,|-\rangle\}$, anyone of them is sufficient to
assist Diana to regenerate the original state of particle $S$ on
particle $D$. This result implies that if we choose Diana as the
receiver in advance, our scheme reduces to a controlled
teleportation scheme \cite{58PRA4394}. It is worth pointing out that
the controlled teleportation schemes with GHZ states
\cite{58PRA4394} are very fragile to the loss of the supervisors'
measurement information. That is, if Bob does not successfully
receive the single-particle measurement outcome of anyone of
supervisors, he cannot recover Alice's original state. In contrast,
our scheme can endure the loss of the measurement information of one
of the supervisors (Bob and Charlie). The controlled teleportation
scheme of Ref.~\cite{8QIP319} also has such a feature, but in which
the teleportation fidelity is less than one.

If both Bob and Charlie choose the measurement basis
$\{|0\rangle,|1\rangle\}$, the transformations that Diana should
perform in order to reconstruct Alice's qubit, up to a global phase,
are
\begin{eqnarray}
 |\Psi^{+}_{SA}\rangle|q_Bq_C\rangle\rightarrow I, & & |\Phi^{+}_{SA}\rangle|q_Bq_C\rangle\rightarrow \sigma_x, \nonumber\\
 |\Psi^{+}_{SA}\rangle|q_B\bar{q}_C\rangle\rightarrow \sigma_x\sigma_z, && |\Phi^{+}_{SA}\rangle|q_B\bar{q}_C\rangle\rightarrow \sigma_z, \nonumber\\
 |\Psi^{-}_{SA}\rangle|q_Bq_C\rangle\rightarrow \sigma_z, & &|\Phi^{-}_{SA}\rangle|q_Bq_C\rangle\rightarrow \sigma_x\sigma_z, \nonumber\\
 |\Psi^{-}_{SA}\rangle|q_B\bar{q}_C\rangle\rightarrow \sigma_x, && |\Phi^{-}_{SA}\rangle|q_B\bar{q}_C\rangle\rightarrow I, \nonumber\\
\end{eqnarray}
where $q\in \{0,1\}$ and $\bar{q}$ is the counterpart of the binary
number $q$. As to the case where Bob or Charlie choose the
measurement basis $\{|+\rangle,|-\rangle\}$, the transformations
that Diana should perform in order to reconstruct Alice's qubit, up
to an overall sign, are
\begin{eqnarray}
 |\Psi^{+}_{SA}\rangle|+_{B(C)}\rangle\rightarrow \sigma_x H, & & |\Phi^{+}_{SA}\rangle|+_{B(C)}\rangle\rightarrow H, \nonumber\\
 |\Psi^{+}_{SA}\rangle|-_{B(C)}\rangle\rightarrow \sigma_z H, && |\Phi^{+}_{SA}\rangle|-_{B(C)}\rangle\rightarrow \sigma_x\sigma_z H, \nonumber\\
 |\Psi^{-}_{SA}\rangle|+_{B(C)}\rangle\rightarrow \sigma_x\sigma_z H, & &|\Phi^{-}_{SA}\rangle|+_{B(C)}\rangle\rightarrow \sigma_z H, \nonumber\\
 |\Psi^{-}_{SA}\rangle|-_{B(C)}\rangle\rightarrow  H, && |\Phi^{-}_{SA}\rangle|-_{B(C)}\rangle\rightarrow \sigma_x H, \nonumber\\
\end{eqnarray}
where $H$ is the Hardamard transformation given by
\begin{equation}
 H=\frac{1}{\sqrt{2}}\left(
 \begin{array}{cc}
   1 & 1\\
   1 & -1\\
 \end{array}
 \right),
\end{equation}
which functions as $H|0\rangle=|+\rangle$ and
$H|1\rangle=|-\rangle$.

In conclusion, we have proposed a scheme for hierarchical QIS, where
the authorities of the three agents, i.e., Bob, Charlie, and Diana,
for getting the quantum secret are hierarchized. That is, Diana has
a larger authority than Bob and Charlie to possess the quantum
secret. The security checking for the quantum channel is the same
that of Ref.~\cite{282OC2457}. Our scheme can also be modified to
implement controlled teleportation against uncooperation of part of
supervisors. Recently, different methods for preparing the state
$|\chi\rangle$ have been proposed \cite{78PRA024301,282OC1052}.
These achievements may contribute to our scheme in physical
realization. In the future, one can generalize the idea to a more
general case where more than three agents are involved.

The hierarchical QIS may be very interesting in view of the
reliability of the agents in quantum communication and the access
controlling in architecture of quantum computer, and should be more
useful than the symmetric QIS in practice. Let us take a simple
example that a dealer in Berlin wants to have an action taken on her
behalf in Beijing. She has many agents who can carry it out for her,
but she knows that some of them are dishonest and does not know whom
they are. She cannot simply send a message to one of them, because
the dishonest ones will try to sabotage the action, but she knows
that if all of them carry it out together, the honest ones will keep
the dishonest ones from doing any damage. Then she can encode the
message in a quantum state (quantum secret) and distribute it among
them through the generalized teleportation protocol discussed above.
The agent who is the most reliable will be distributed a larger part
of information. As a consequence, the most reliable agent can
recover the secret with the cooperation of subset of the other ones,
but the other ones cannot get the secret without the participation
of the most reliable one.

\begin{acknowledgements}
  This work was supported by the Natural Science Foundation of
Hunan Province of China (Grant No. 06JJ50015).
\end{acknowledgements}

\end{document}